# Scanning SQUID characterization of extremely overdoped La$_{2-x}$Sr$_x$CuO$_4$

Chloe Herrera[1,2], Jacob Franklin[1], Ivan Božović[3,4], Xi He[3] and Ilya Sochnikov[1,5]

[1]Physics Department, University of Connecticut, Storrs CT 06269, USA
[2]Lux Research, Boston MA 02110, USA
[3]Brookhaven National Laboratory, Upton, New York 11973-5000, USA
[4]Department of Chemistry, Yale University, New Haven CT 06520, USA
[5]Institute of Material Science, University of Connecticut, Storrs CT 06269, USA

**Recently, advances in film synthesis methods have enabled a study of extremely overdoped La$_{2-x}$Sr$_x$CuO$_4$. This has revealed a surprising behavior of the superfluid density as a function of doping and temperature, the explanation of which is vividly debated. One popular class of models posits electronic phase separation, where the superconducting phase fraction decreases with doping, while some competing phase (e.g. ferromagnetic) progressively takes over. A problem with this scenario is that all the way up to the dome edge the superconducting transition remains sharp, according to mutual inductance measurements. However, the physically relevant scale is the Pearl penetration depth, $\Lambda_P$, and this technique probes the sample on a length scale $L$ that is much larger than $\Lambda_P$. In the present paper, we use local scanning SQUID measurements that probe the susceptibility of the sample on the scale $L \ll \Lambda_P$. Our SQUID maps show uniform landscapes of susceptibility and excellent overall agreement of the local penetration depth data with the bulk measurements. These results contribute an important piece to the puzzle of how high-temperature superconductivity vanishes on the overdoped side of the cuprates phase diagram.**

## I. INTRODUCTION

Prior to the discovery of cuprate superconductors with high critical temperatures ($T_c$) [1,2], it was widely believed that the theory of Bardeen, Cooper and Schrieffer (BCS) [3] should be applicable to every superconducting material. In BCS theory, electron pairing arises from electron-phonon interactions; electrons of opposite spins exchange phonons, which ultimately produces an attraction between the two electrons which form a Cooper pair. However, cuprates are not ordinary metals; they harbor strong electron correlations and are usually modeled as doped antiferromagnetic Mott insulators. In the temperature versus doping phase diagram, $T_c$ traces a characteristic dome-shaped curve. So far, most of the efforts were focused on the underdoped side of the phase diagram, where a pseudogap and a rich variety of other orders and phases are observed [2]. Only recently has the attention shifted to the overdoped side, enabled by advances in the synthesis of high-$T_c$ cuprate films, La$_{2-x}$Sr$_x$CuO$_4$ (LSCO) in particular [4]. Bulk crystals with such high doping levels are very hard to obtain, if they can be obtained at all.

A systematic study of overdoped LSCO samples may provide insight into the underlying cause(s) of the decrease and disappearance of superconductivity with doping, and thus enable a better understanding of the high-$T_c$ superconductivity mechanism in general. In the past, it was usually assumed that the physics of overdoped cuprates should conform to the standard BCS theory of the superconducting state and the Fermi liquid theory of the normal-metal state. However, the applicability of this conventional picture has been challenged recently, as the closer scrutiny of the overdoped region has generated several perplexing experimental discoveries [5]. One of the key discrepancies is in the temperature and doping dependence of superfluid stiffness $N_s$, a quantity inversely proportional to the square of $\lambda_L$, the magnetic field (London) penetration depth in the overdoped regime [6–14]. Contrary to the expectations, the $N_s(T)$ dependence was found to be essentially linear, from the lowest temperature all the way up to $T_c$. Even more perplexing, as the doping level is increased, $N_{s0} \equiv N_s(T \rightarrow 0)$ tracks $T_c$, i.e., it keeps decreasing, even though the overall carrier density and the materials' conductivity keep increasing.

With this in mind, it has been proposed that perhaps some sort of electronic phase separation occurs on the overdoped side. The samples are postulated to be very inhomogeneous, with one phase superconducting and the other not, due to some competing order, such as ferromagnetism [15]. As one approaches the quantum critical point at which superconductivity disappears, the superconducting fraction decreases, and the competing phase grows. In accord with this proposal, a systematic study of overdoped LSCO films by THz spectroscopy has shown that indeed the superfluid δ-function coexists with a Drude-like response from normal carriers, even at the lowest temperature (1.5 K) accessible in that experiment [16]. As the doping is increased and superfluid density decreases, the normal component spectral weight increases, so that the Ferrell-Glover-Tinkham sum rule indeed remains satisfied.

One problem with such phase-separation scenarios, otherwise quite plausible, is that a dedicated search for such a competing order has so far failed to detect any [14]. Another is that, according to mutual inductance measurements, the superconducting transition in MBE-grown LSCO films remains very sharp all the way to the dome edge. In the best films, one can put an upper limit on any variations in $T_c$ to a fraction of a degree Kelvin. However, the mutual inductance technique probes the sample on a length scale $L$ of several millimeters. For screening of the magnetic field by a thin superconducting film, the physically relevant, intrinsic length scale is the so-called Pearl penetration depth, $\Lambda_P = 2\lambda_L^2/d$, where $\lambda_L$ is the usual London penetration depth and $d$ is the film thickness. Since $\lambda_L \sim$ 300-600 nm in the OD regime and at the lowest temperature we measure, while the typical thickness of our films is $d \sim$20-30 nm, we infer that 6 μm $< \Lambda_P <$ 36 μm. Comparing the two length scales, the mutual inductance technique always operates in the regime $L >> \Lambda_P$. This, in principle, leaves some possibility of local inhomogeneity to have been left undetected.

In this work, we aim to quantitatively determine penetration depth locally, scan across the sample to image with a micrometer resolution what happens in the material at temperatures near $T_c$, and check for spatial inhomogeneities in $T_c$ and $N_s(T)$. We present a comparison between previous global measurements and our present local scanning Superconducting Quantum Interference Device (SQUID) measurements of susceptibility. The two data sets are found to be in excellent agreement, implying that gross disorder cannot explain the exotic behavior of the superfluid density.

## II. EXPERIMENT

All scanning SQUID experiments were done at the University of Connecticut using the Montana instruments Fusion 2 Cryostation, with a home-built microscope (FIG. 1). Scanning SQUID microscopy (FIG. 1) can resolve micrometer-scale variations in both magnetic and susceptibility signatures in the material under study. The SQUID used in this work is a gradiometric susceptometer, which can be thought of as a miniaturized version of the mutual-inductance setup in the reflection geometry [17,18]. The key difference with most mutual inductance setups is in the dimensions of our SQUID device, which is over two orders of magnitude smaller than the size of inductance coils used in global susceptibility measurements [19], to be precise 1.8 mm coils in bulk experiments versus 14 and 7 $\mu m$ field coil and pickup loops in the SQUID sensor used here. Scanning the SQUID parallel to the film surface thus allows for micrometer-scale imaging. A small local AC field from the SQUID field coil induces a response from the sample, and the SQUID measures that response via a small pick-up loop (FIG. 1). Gradiometric modulation coils positioned at the center of the device couple to the response, and we measure the modulation current in well-calibrated flux units via a feedback loop amplifier. In order to collect raster images, a piezo-scanner moves in a line-by-line fashion to record the magnetic susceptibility and the dc flux distribution in a rectangular scan area [20–23]. The SQUID microscope also measures susceptibility approach curves by moving towards the sample while keeping the lateral coordinates fixed, which allows us to determine the local magnetic penetration depth. All measurements can be done without any special preparation of the material, such as lithography or soldering electrical contacts.

High-quality, single-crystal LSCO films were synthesized at the Brookhaven National Laboratory using atomic layer-by-layer molecular beam epitaxy (ALL-MBE) [24,25]. The custom built ALL-MBE system has the following key components: (1) 16 elemental sources that can be individually controlled via computer-commanded pneumatic shutter mechanisms; (2) a pure ozone supply, with a water-cooled delivery nozzle; (3) a 6-degrees of freedom substrate positioner and heater; (4) a reflection high energy electron diffraction (RHEED) system, which enables monitoring the surface crystal structure and morphology in real time during the film synthesis; (6) a system for measurements of metal source fluxes [26] (FIG. 1).

The LSCO films studied here were deposited on LaSrAlO$_4$ (LSAO) substrates epitaxially polished with the surface perpendicular to the [001] crystallographic direction. Each of the two LSCO films under study consists of an active (superconducting) layer, 25 nm (19 unit cells, 1.32 nm each) thick, which is protected on both sides with a 13 nm (5 unit cells) thick layer of La$_{1.60}$Sr$_{0.40}$CuO$_4$, which is metallic but not superconducting (FIG. 1). These protective layers ensure excellent film stability and immunity to exposure to atmosphere.

The nominal composition of the active layers in the two films under study is La$_{2-x}$Sr$_x$CuO$_4$ with $x$ = 0.32 and $x$ = 0.33, respectively. We emphasize that this x, the Sr concentration, should not be conflated with the actual density $p$ of mobile charge carriers in the film. The two are different because of the presence of an unknown concentration and distribution of oxygen vacancies. Local defects that may cause some electron localization, etc., are also possible. In fact, we are unaware of any accurate and reliable method of measuring $p$, and so it remains unknown. In the vast literature on high-$T_c$ cuprates, it has been an accepted practice to infer $p$ from the measured value of $T_c$, using the following conversion: $T_c = B(p - p_{c1})(p_{c2} - p)$ where $p_{c1} = 0.06$, $p_{c2} = 0.26$, and (for LSCO, specifically) $B = 4.15 \times 10^3$ K. In both LSCO films studied here, the mutual inductance and SQUID measurements show $T_c \approx 12$ K. If we use the above conversion, the hole concentration determined in this way would be $p \approx 0.25$, indeed very close to the dome edge at $p_{c2} = 0.26$. In any case, we are way on the overdoped side, which is of focal interest here. Since the overall superfluid density scales with $T_c$, it is 3.5 times reduced compared to that in the optimally doped LSCO, while the total carrier density is higher. In a naïve spatial-phase-separation model, one would expect most of the of film — more than three-quarters — not to be superconducting, even at the lowest temperature, and presumably we should not fail to observe that clearly.

In what follows, we present the acquired SQUID microscopy data including scan images and susceptibility approach curves.

## III. RESULTS

The SQUID microscope scans are presented as images (FIG. 2 and FIG. 3) over the temperature range 6 K < $T$ < 14 K. Susceptibility approach curves are presented with a full temperature series on a single location of the sample, with two locations displayed for each sample (FIG. 4 and FIG. 5).

For scan images (FIG. 2 and FIG. 3), the color-bar intensity corresponds to susceptibility signal. Darker blue indicates a signal tending toward the superconducting diamagnetic state while yellow indicates a paramagnetic direction. For each doping level, scans were taken from the lowest temperature on the sample holder up to a few Kelvin above $T_c$, determined by observing at which point the features in the scan images disappear. The color bars are chosen to have an upper limit near the average susceptibility over the scan area and a fixed lower limit near zero (after background signal subtraction). No software filters or image processing were applied except for standard background subtraction and removal of a few occasional scan-lines that occurred when the SQUID lost feedback control due to external electromagnetic noise such as cell phone signals. Otherwise, these are essentially raw data (FIG. 2 and FIG. 3), which is important for proper analysis.

The temperature dependence of both LSCO films reveal homogenous susceptibility, at the level of a few percent. This translates to similar characteristic homogeneity in the penetration depth and the superfluid density. The transitions to the normal state are well-defined and are very sharp with a width of about 0.5 K. In this respect, the present films rival the best superconducting crystals [27].

FIG. 4 shows typical dependence of susceptibility signal as a function of voltage on the $Z$ piezo, which is proportional to the distance between the sample and the SQUID. As the distance between the sample and SQUID decreases, the susceptibility diverges. Each approach curve corresponds to one temperature in the range from less than 6 K up to 14.5 K, well above $T_c$. The susceptibility approach curves diminish as the temperature is increased. When the material stops being superconducting the approach curve flattens out within the noise level.

The susceptibility approach curve is a continuous measurement of magnetic susceptibility from the sample as it moves closer to the SQUID. From this approach curve, the Pearl length $\Lambda_P$ can be determined. Using a model developed by Kogan [28] and approximated for a thin diamagnet by Kirtley et al. [18], we find the $\Lambda_P$ for the two overdoped LSCO films under study, as a function of temperature. The model assumes that the SQUID pick-up loop and the field coil are circle-shaped, the leads are infinitely thin, and the penetration depth $\lambda_L$ is much larger than film thickness $d$. In this limit, the model describes the susceptibility approach curves as $\phi(z) = \frac{\phi_s a}{\Lambda_p}\left(1 - \frac{2\bar{z}}{\sqrt{1+4\bar{z}^2}}\right)$, where $\phi(z)$ is the measured susceptibility signal, $\phi_s = 1711\ \Phi_0/A$ is the mutual inductance between the field coil and the pickup loop determined numerically for a specific sensor used, $a = 7\ \mu m$ is the radius of the field coil, and $\bar{z} = z/a$ is the distance between the SQUID and the sample, normalized by $a$. Both the susceptibility images and the susceptibility approach traces are acquired with the field coil current of $\sim 15\ \mu A$ at a typical frequency ~1700 Hz, corresponding approximately to ~10 mG amplitude at the center of the coil when it is at the closest position to the sample.

Determining $\bar{z} = (z_0 + kV)/a$ takes two considerations: a constant offset between the SQUID and the sample due to the SQUID alignment at a finite angle, $z_0$, and a piezo calibration factor $k$ to convert the voltage on the piezo, $V$, during approach to micrometers. Utilizing the gradiometric design of the SQUID chip necessitates a small angular alignment. The alignment angle in our cool-down cycles was $\sim 2$ - $4°$ between the SQUID chip and the sample surface, as determined optically at cryogenic temperatures. This creates a distance $z_0$ between the SQUID and sample when the capacitance measurements indicate a 'touchdown' (contact between the SQUID and the sample). As one can see in the expression for $\phi(z)$ for thin films, the penetration depth $\Lambda_P$ is simply a factor dividing the $\bar{z}$-dependent term. We estimate $z_0 = 2.5, 2.3, 2.1, 1.8\ \mu m$, and $k = 7.5, 6.6, 7.5,$ $7.5\ \mu m/V$ for the four datasets shown in the text and the supplementary information (in the order of appearance, starting from FIG. 4). The systematic uncertainty in $z_0$ prevails in SQUID measurements as the source of the error, in general. We estimate that uncertainty to be about $\pm 0.5$ $\mu m$ leading to the error bars shown in FIG. 5. This statement holds unless the signals are small and random errors come into play, which is not the case in our measurements for most temperatures, except for those very close to $T_c$. With these parameters we fit the above model for $\phi(z)$ to the approach traces of susceptibility, which gives us the value of the Pearl penetration depth, $\Lambda_P$. The latter is translated to the superfluid phase stiffness as explained in the following.

## IV. DISCUSSION

The Pearl depth $\Lambda_P \propto 1/N_s$ is related to the phase stiffness $\rho_s = 2A/d\Lambda_P$, where $A = \hbar^2 c_0/8\mu_0 k_B e^2$, $\hbar$ is the reduced Planck constant, $c_0 = 1.32$ nm is the unit cell height, $\mu_0$ is the vacuum permeability, $k_B$ is the Boltzmann constant, and $e$ is the electron charge [4]. Note, this definition is independent of the film thickness consistent with our previous bulk measurements [4] that showed no dependence of the superfluid phase stiffness on the film thickness. FIG. 5 shows that the $\rho_s(T)$ dependence extracted from the SQUID approach data is indeed linear in $T$, exactly as observed in the global measurements [5]. The lack of gross obvious inhomogeneities besides some minor noise and artifacts in our scans indicate that the origin of this unusual T-linear dependence of $\rho_s$ is intrinsic.

Several theoretical explanation of the unusual temperature and doping dependence of $\rho_s(T,p)$ in LSCO and other cuprates have been proposed already [29–42]. The most conventional approach is based on the standard BCS theory of "dirty" superconductors and the assumption that the demise of $T_c$ and of $\rho_{s0}$ with doping is due to the increasing disorder caused by the increasing density of random Sr dopants [31,32]. However, this model runs into a lot of difficulties and contradictions. One conceptual problem is that, experimentally, the coherence length is much smaller than the mean free path right above $T_c$. Hence, if the standard BCS theory applied, we should be in the "clean-BCS" limit. In principle, this would be consistent with the observed temperature dependence of $\rho_s(T,p)$. But it is in contradiction with the facts that most carriers are not in superfluid even when $T{\rightarrow}0$, and that $\rho_{s0}$ decreases and vanishes as $p \rightarrow p_{c2}$, while the normal carrier density keeps increasing smoothly. This problem was clearly spelled out early on [29] and directly corroborated subsequently by THz spectroscopy [16]. Additional quantitative and qualitative discrepancies have been pointed out in [35]. For these reasons, several other models that go beyond the conventional have been proposed, emphasizing the role of Mott physics (i.e., strong electron correlations) [37,40], strong phase fluctuations [16,33,39], strong pairing interaction that causes formation of small (non-overlapping) preformed pairs that undergo Bose-Einstein condensation [36], etc., but so far without consensus. A detailed discussion of all the arguments pro and con of each of these models is outside of the scope of present work.

One inference that seems to be very robust and non-controversial is that the measured superfluid density decreases and becomes very low in the highly-overdoped region. When expressed as the superfluid stiffness in units of Kelvin, this can be interpreted as the characteristic temperature at which thermal phase fluctuations become strong enough to destroy global phase coherence. This characteristic temperature is roughly equal to $T_c$, implying that $T_c$ is in fact controlled by phase fluctuations. We believe that this important problem is still open and calls for more theoretical work.

## V. SUMMARY AND CONCLUSIONS

To summarize, we found excellent agreement between the local SQUID measurements and the previous bulk measurements of the penetration depth. This confirmation allows for a closer inspection on the quantitative information garnered from local measurements. Scanning SQUID susceptometry provides a micrometer-scale spatial depiction of the development of superconductivity that cannot be achieved via other methods like the bulk mutual inductance technique. Using SQUIDs to find the penetration depth can be a reliable method, once the fitting parameters are reasonably constrained. SQUIDs are cited as being non-invasive and we note that it is only true when the samples are measured very carefully, and even a smallest bit of contact between the SQUID and sample may alter the results. When done carefully to avoid interference with the device, it's possible to gain new insight on micrometer-scale developments of magnetic properties of novel quantum materials.

## ACKNOWLEDGEMENTS

Work at Brookhaven National Laboratory was supported by the DOE, Basic Energy Sciences, Materials Sciences and Engineering Division. X. H. is supported by the Gordon and Betty Moore Foundation's EPiQS Initiative through grant GBMF9074. Work at UCONN was in part supported by the US DOD.

## References

[1] P. A. Lee, N. Nagaosa, and X.-G. Wen, *Doping a Mott Insulator: Physics of High-Temperature Superconductivity*, Rev. Mod. Phys. **78**, 17 (2006).

[2] B. Keimer, S. A. Kivelson, M. R. Norman, S. Uchida, and J. Zaanen, *From Quantum Matter to High-Temperature Superconductivity in Copper Oxides*, Nature **518**, 179 (2015).

[3] J. Bardeen, L. N. Cooper, and J. R. Schrieffer, *Theory of Superconductivity*, Phys Rev **108**, 1175 (1957).

[4] I. Božović, X. He, J. Wu, and A. T. Bollinger, *Dependence of the Critical Temperature in Overdoped Copper Oxides on Superfluid Density*, Nature **536**, 309 (2016).

[5] I. Božović, X. He, J. Wu, and A. T. Bollinger, *The Demise of Superfluid Density in Overdoped La$_{2-x}$Sr$_x$CuO$_4$ Films Grown by Molecular Beam Epitaxy*, J. Supercond. Nov. Magn. **30**, 1345 (2017).

[6] Y. J. Uemura, G. M. Luke, B. J. Sternlieb, J. H. Brewer, J. F. Carolan, W. N. Hardy, R. Kadono, J. R. Kempton, R. F. Kiefl, S. R. Kreitzman, P. Mulhern, T. M. Riseman, D. Ll. Williams, B. X. Yang, S. Uchida, H. Takagi, J. Gopalakrishnan, A. W. Sleight, M. A. Subramanian, C. L. Chien, M. Z. Cieplak, G. Xiao, V. Y. Lee, B. W. Statt, C. E. Stronach, W. J. Kossler, and X. H. Yu, *Universal Correlations between T$_c$ and n$_s$/m$^*$ (Carrier Density over Effective Mass) in High-T$_c$ Cuprate Superconductors*, Phys. Rev. Lett. **62**, 2317 (1989).

[7] Y. J. Uemura, A. Keren, L. P. Le, G. M. Luke, W. D. Wu, Y. Kubo, T. Manako, Y. Shimakawa, M. Subramanian, J. L. Cobb, and J. T. Markert, *Magnetic-Field Penetration Depth in Tl$_2$Ba$_2$CuO$_{6+\delta}$ in the Overdoped Regime*, Nature **364**, 605 (1993).

[8] C. Niedermayer, C. Bernhard, U. Binninger, H. Glückler, J. L. Tallon, E. J. Ansaldo, and J. I. Budnick, *Muon Spin Rotation Study of the Correlation between T$_c$ and n$_s$/m$^*$ in Overdoped T$_2$Ba$_2$CuO$_{6+\delta}$*, Phys. Rev. Lett. **71**, 1764 (1993).

[9] C. Bernhard, Ch. Niedermayer, U. Binninger, A. Hofer, Ch. Wenger, J. L. Tallon, G. V. M. Williams, E. J. Ansaldo, J. I. Budnick, C. E. Stronach, D. R. Noakes, and M. A. Blankson-Mills, *Magnetic Penetration Depth and Condensate Density of Cuprate High-T$_c$ Superconductors Determined by Muon-Spin-Rotation Experiments*, Phys. Rev. B **52**, 10488 (1995).

[10] C. Panagopoulos, T. Xiang, W. Anukool, J. R. Cooper, Y. S. Wang, and C. W. Chu, *Superfluid Response in Monolayer High-T$_c$ Cuprates*, Phys. Rev. B **67**, 220502 (2003).

[11] J.-P. Locquet, Y. Jaccard, A. Cretton, E. J. Williams, F. Arrouy, E. Mächler, T. Schneider, O. Fischer, and P. Martinoli, *Variation of the In-Plane Penetration Depth λ$_{ab}$ as a Function of Doping in La$_{2-x}$Sr$_x$CuO$_{4\pm\delta}$ Thin Films on SrTiO$_3$: Implications for the Overdoped State*, Phys. Rev. B **54**, 7481 (1996).

[12] T. R. Lemberger, I. Hetel, A. Tsukada, M. Naito, and M. Randeria, *Superconductor-to-Metal Quantum Phase Transition in Overdoped La$_{2-x}$Sr$_x$CuO$_4$*, Phys. Rev. B **83**, 140507 (2011).

[13] P. M. C. Rourke, I. Mouzopoulou, X. Xu, C. Panagopoulos, Y. Wang, B. Vignolle, C. Proust, E. V. Kurganova, U. Zeitler, Y. Tanabe, T. Adachi, Y. Koike, and N. E. Hussey, *Phase-Fluctuating Superconductivity in Overdoped La$_{2-x}$Sr$_x$CuO$_4$*, Nat. Phys. **7**, 455 (2011).

[14] I. Božović, X. He, J. Wu, and A. T. Bollinger, *The Vanishing Superfluid Density in Cuprates—and Why It Matters*, J. Supercond. Nov. Magn. **31**, 2683 (2018).

[15] J. Wu, V. Lauter, H. Ambaye, X. He, and I. Božović, *Search for Ferromagnetic Order in Overdoped Copper-Oxide Superconductors*, Sci. Rep. **7**, 45896 (2017).

[16] F. Mahmood, X. He, I. Božović, and N. P. Armitage, *Locating the Missing Superconducting Electrons in the Overdoped Cuprates La$_{2-x}$Sr$_x$CuO$_4$*, Phys. Rev. Lett. **122**, 027003 (2019).

[17] M. E. Huber, N. C. Koshnick, H. Bluhm, L. J. Archuleta, T. Azua, P. G. Björnsson, B. W. Gardner, S. T. Halloran, E. A. Lucero, and K. A. Moler, *Gradiometric Micro-SQUID Susceptometer for Scanning Measurements of Mesoscopic Samples*, Rev. Sci. Instrum. **79**, 053704 (2008).

[18] J. R. Kirtley, B. Kalisky, J. A. Bert, C. Bell, M. Kim, Y. Hikita, H. Y. Hwang, J. H. Ngai, Y. Segal, F. J. Walker, C. H. Ahn, and K. A. Moler, *Scanning SQUID Susceptometry of a Paramagnetic Superconductor*, Phys. Rev. B **85**, 224518 (2012).

[19] X. He, A. Gozar, R. Sundling, and I. Božović, *High-Precision Measurement of Magnetic Penetration Depth in Superconducting Films*, Rev. Sci. Instrum. **87**, 113903 (2016).

[20] B. Kalisky, J. R. Kirtley, J. G. Analytis, J.-H. Chu, A. Vailionis, I. R. Fisher, and K. A. Moler, *Stripes of Increased Diamagnetic Susceptibility in Underdoped Superconducting Ba(Fe$_{1-x}$Co$_x$)$_2$As$_2$ Single Crystals: Evidence for an Enhanced Superfluid Density at Twin Boundaries*, Phys. Rev. B **81**, 184513 (2010).

[21] I. Sochnikov, A. J. Bestwick, J. R. Williams, T. M. Lippman, I. R. Fisher, D. Goldhaber-Gordon, J. R. Kirtley, and K. A. Moler, *Direct Measurement of Current-Phase Relations in Superconductor/Topological Insulator/Superconductor Junctions*, Nano Lett. **13**, 3086 (2013).

[22] I. Sochnikov, L. Maier, C. A. Watson, J. R. Kirtley, C. Gould, G. Tkachov, E. M. Hankiewicz, C. Brüne, H. Buhmann, L. W. Molenkamp, and K. A. Moler, *Nonsinusoidal Current-Phase Relationship in Josephson Junctions from the 3D Topological Insulator HgTe*, Phys. Rev. Lett. **114**, 066801 (2015).

[23] H.-Y. Yang, B. Singh, J. Gaudet, B. Lu, C.-Y. Huang, W.-C. Chiu, S.-M. Huang, B. Wang, F. Bahrami, B. Xu, J. Franklin, I. Sochnikov, D. E. Graf, G. Xu, Y. Zhao, C. M. Hoffman, H. Lin, D. H. Torchinsky, C. L. Broholm, A. Bansil, and F. Tafti, *A New Noncollinear Ferromagnetic Weyl Semimetal with Anisotropic Anomalous Hall Effect*, ArXiv:2006.07943 (2020).

[24] I. Bozovic, *Atomic Layer Engineering of Superconducting Oxides: Yesterday, Today, Tomorrow*, IEEE Trans Appl Supercond **11**, 2686 (2001).

[25] G. Logvenov, A. Gozar, and I. Bozovic, *High-Temperature Superconductivity in a Single Copper-Oxygen Plane*, Science **326**, 699 (2009).

[26] J. N. Eckstein and I. Bozovic, *High-Temperature Superconducting Multilayers and Heterostructures Grown by Atomic Layer-By-Layer Molecular Beam Epitaxy*, Annu. Rev. Mater. Sci. **25**, 679 (1995).

[27] T. M. Lippman, B. Kalisky, H. Kim, M. A. Tanatar, S. L. Bud'ko, P. C. Canfield, R. Prozorov, and K. A. Moler, *Agreement between Local and Global Measurements of the London Penetration Depth*, Phys. C Supercond. **483**, 91 (2012).

[28] V. G. Kogan, *Meissner Response of Anisotropic Superconductors*, Phys. Rev. B **68**, 104511 (2003).

[29] J. Zaanen, *Superconducting Electrons Go Missing*, Nature **536**, 282 (2016).

[30] E. A. Pashitskii, *The Critical Temperature as a Function of the Number of Cooper Pairs, and the Superconductivity Mechanism in a Layered LaSrCuO Crystal*, Low Temp. Phys. **42**, 1184 (2016).

[31] N. R. Lee-Hone, J. S. Dodge, and D. M. Broun, *Disorder and Superfluid Density in Overdoped Cuprate Superconductors*, Phys. Rev. B **96**, 024501 (2017).

[32] V. R. Shaginyan, V. A. Stephanovich, A. Z. Msezane, G. S. Japaridze, and K. G. Popov, *The Influence of Topological Phase Transition on the Superfluid Density of Overdoped Copper Oxides*, Phys. Chem. Chem. Phys. **19**, 21964 (2017).

[33] Y. Tao, *BCS Quantum Critical Phenomena*, EPL Europhys. Lett. **118**, 57007 (2017).

[34] N. R. Lee-Hone, V. Mishra, D. M. Broun, and P. J. Hirschfeld, *Optical Conductivity of Overdoped Cuprate Superconductors: Application to La$_{2-x}$Sr$_x$CuO$_4$*, Phys. Rev. B **98**, 054506 (2018).

[35] I. Božović, A. T. Bollinger, J. Wu, and X. He, *Can High-Tc Superconductivity in Cuprates Be Explained by the Conventional BCS Theory?*, Low Temp. Phys. **44**, 519 (2018).

[36] V. Lakhno, *Superconducting Properties of a Nonideal Bipolaron Gas*, Phys. C Supercond. Its Appl. **561**, 1 (2019).

[37] V. A. Khodel, J. W. Clark, and M. V. Zverev, *Impact of Electron-Electron Interactions on the Superfluid Density of Dirty Superconductors*, Phys. Rev. B **99**, 184503 (2019).

[38] D. Pelc, P. Popčević, M. Požek, M. Greven, and N. Barišić, *Unusual Behavior of Cuprates Explained by Heterogeneous Charge Localization*, Sci. Adv. **5**, eaau4538 (2019).

[39] Y. Tao, *Relativistic Ginzburg–Landau Equation: An Investigation for Overdoped Cuprate Films*, Phys. Lett. A **384**, 126636 (2020).

[40] P. W. Phillips, L. Yeo, and E. W. Huang, *Exact Theory for Superconductivity in a Doped Mott Insulator*, Nat. Phys. **16**, 1175 (2020).

[41] Y. Liu, Y. Mou, and S. Feng, *Doping Dependence of Electromagnetic Response in Cuprate Superconductors*, J. Supercond. Nov. Magn. **33**, 69 (2020).

[42] Zi-Xiang Li, S. Kivelson, and D.-H. Lee, *The Superconductor to Metal Transition in Overdoped Cuprates*, ArXiv:2010.06091 (2020).

[43] Q. Lei, M. Golalikhani, B. A. Davidson, G. Liu, D. G. Schlom, Q. Qiao, Y. Zhu, R. U. Chandrasena, W. Yang, A. X. Gray, E. Arenholz, A. K. Farrar, D. A. Tenne, M. Hu, J. Guo, R. K. Singh, and X. Xi, *Constructing Oxide Interfaces and Heterostructures by Atomic Layer-by-Layer Laser Molecular Beam Epitaxy*, Npj Quantum Mater. **2**, 10 (2017).

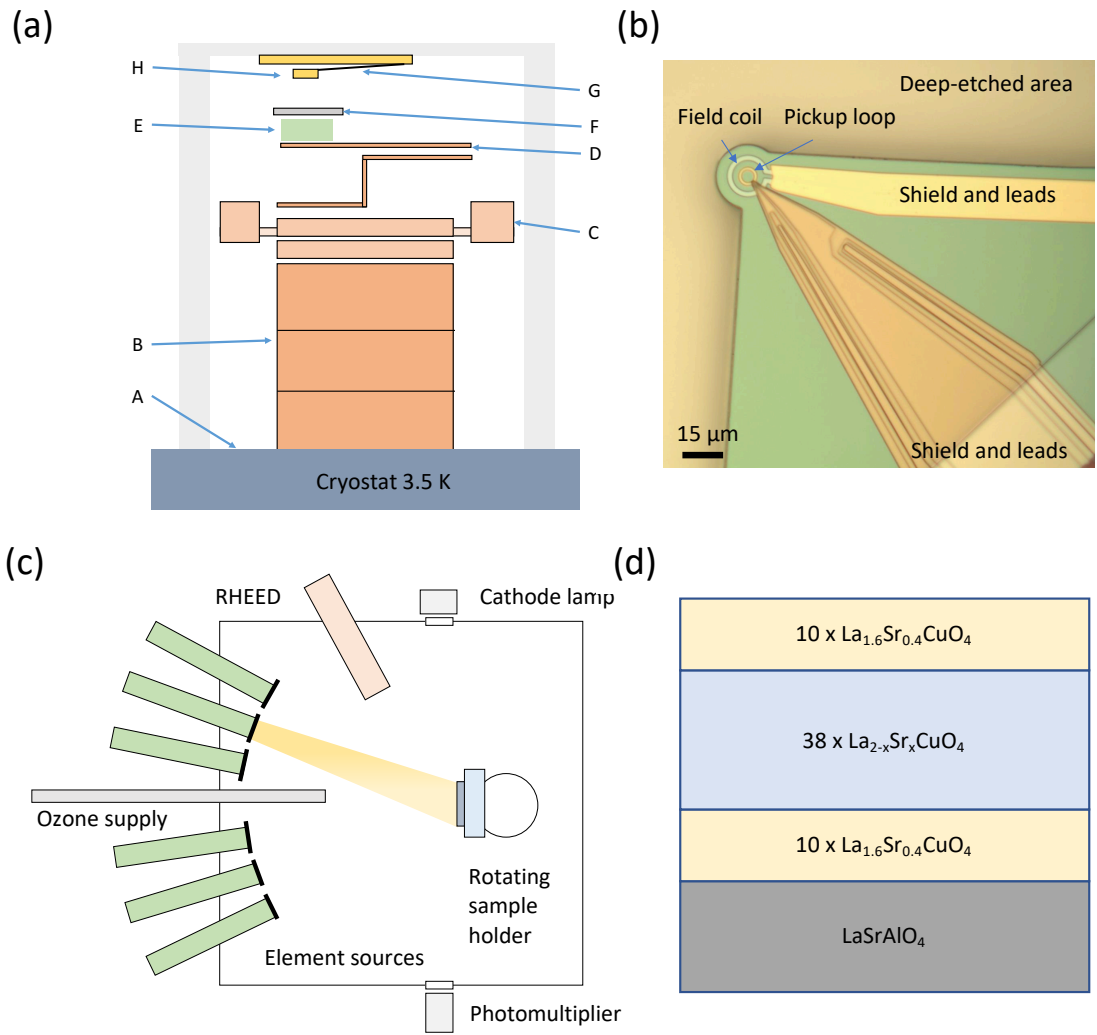

FIG. 1. (a) Schematic of scanning SQUID microscope inside cryogenic chamber. 'A' - cryostat platform that typically reaches 3.5 K at its coolest point. 'B' - coarse piezoelectric positioners. Movement in X, Y, and Z directions is possible within the range of 5 mm in any direction. 'C' - fine piezoelectric positioners in the X-Y directions. 'D' - Z piezo held by a copper brace; it moves the sample closer to SQUID. 'E' – a copper sample holder attached to the Z piezo. A thermometer, heater, and thermal anchor to the platform are all attached to the sample holder. 'F' - sample on copper holder thermalized using silver-Apiezon ® grease. 'G' – a copper cantilever that attaches the SQUID to the electrical contact board and acts as one conductive surface for topographic capacitance measurements, with the other capacitor surface embedded in the board. (b) An optical microscope image of the SQUID chip with a close-up of the tip, highlighting the pick-up loop and the field coil. (c) A schematics of a modular ALL-MBE system [26,43] containing various elemental sources, an ozone supply with a leak valve, a rotating substrate positioner with sample heater, a reflection high-energy electron diffraction (RHEED) system to monitor surface crystallography and morphology, a hollow-cathode lamp, and a photomultiplier tube for flux monitoring. Note that the actual ALL-MBE system has 16 elemental sources, each supplied with its own atomic-absorption spectroscopy (AAS) flux monitor. All 16 AAS channels can be operated simultaneously in real time during the film synthesis, without interference. (c) A cross-section of a $La_{2-\delta}Sr_xCuO_4$ film deposited a $LaSrAlO_4$ substrate, with protective metallic (but non-superconducting) $La_{1.60}Sr_{0.40}CuO_4$ buffer and top-cover layers. Multipliers indicate the number of molecular layers.

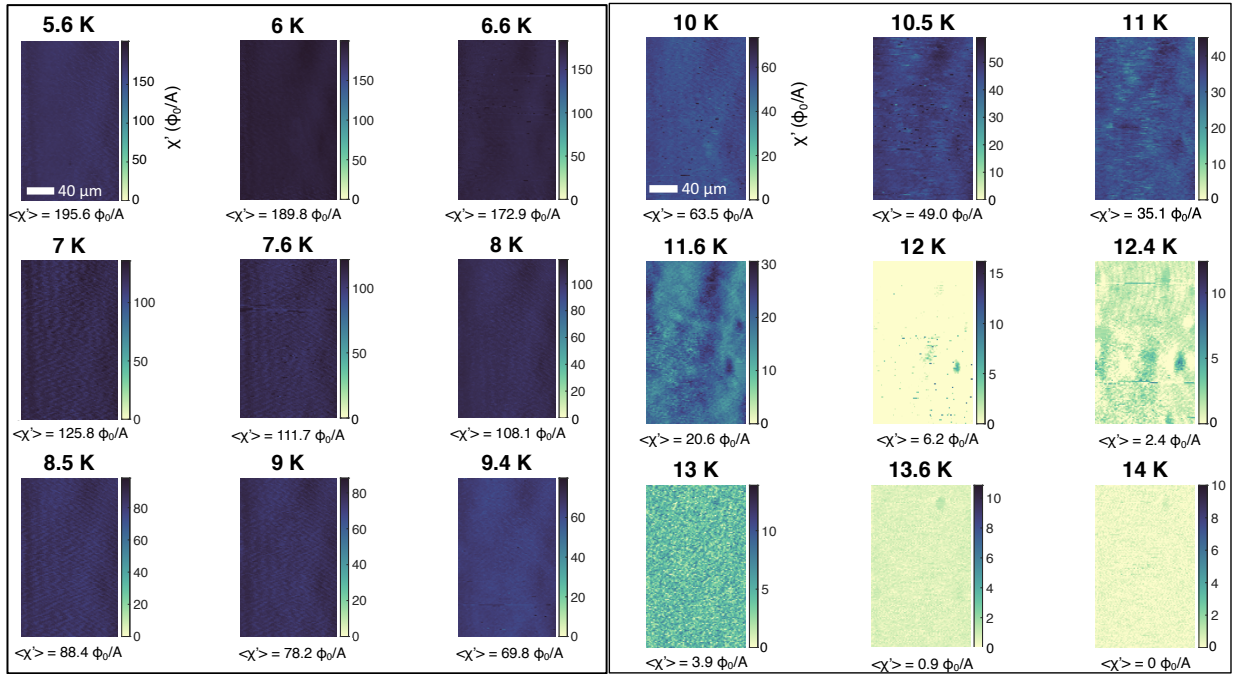

FIG. 2. *Typical scanning SQUID images of overdoped La$_{1.68}$Sr$_{0.32}$CuO$_4$ film synthesized by ALL-MBE. Scan area is 137 × 229 μm². The measurements were performed at a series of temperatures from 5.6 K to 14 K, in order to observe both the superconducting state and the normal state. As temperature is increased, the susceptibility decreases, until it reaches small values indicative of a non-superconducting material. The images show that below T$_c$ the superfluid is homogenous to within few percent level.*

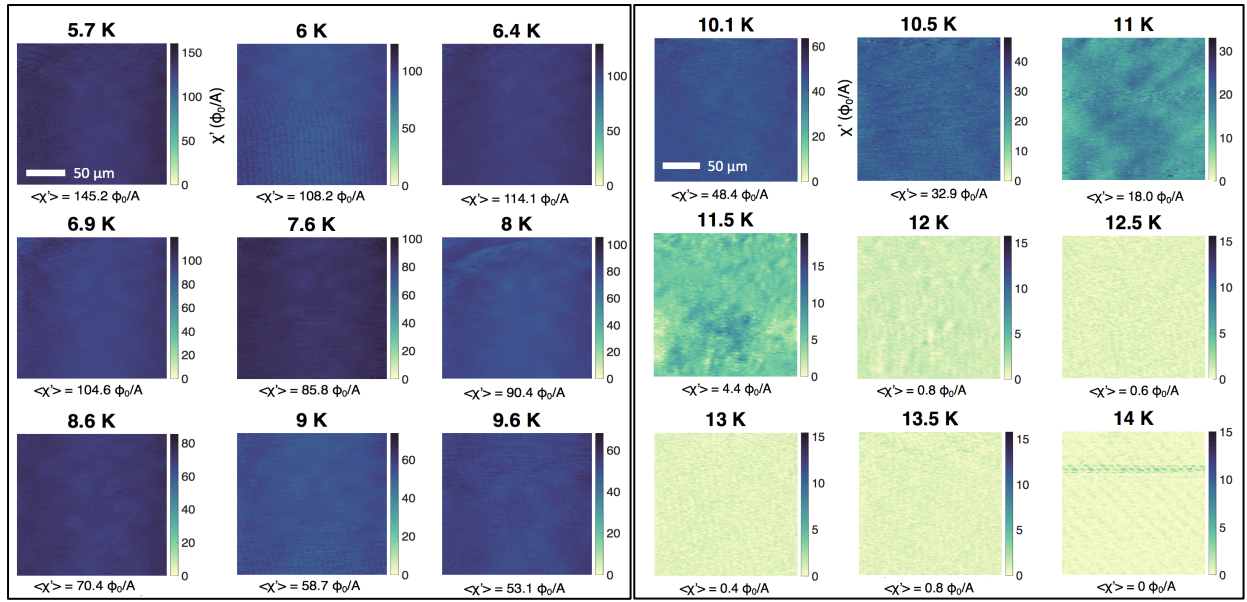

FIG. 3. *Typical ac susceptibility images of overdoped La$_{1.67}$Sr$_{0.33}$CuO$_4$ film synthesized by ALL-MBE. The measurements were performed at a series of temperatures from 5.6 K to 14 K, in order to observe both the superconducting state and the normal state. Scan area is 193 × 181 μm$^2$. As temperature is increased, the susceptibility decreases, until it reaches small values indicative of a non-superconducting material. The images show that below T$_c$ the superfluid is homogenous to within few percent level.*

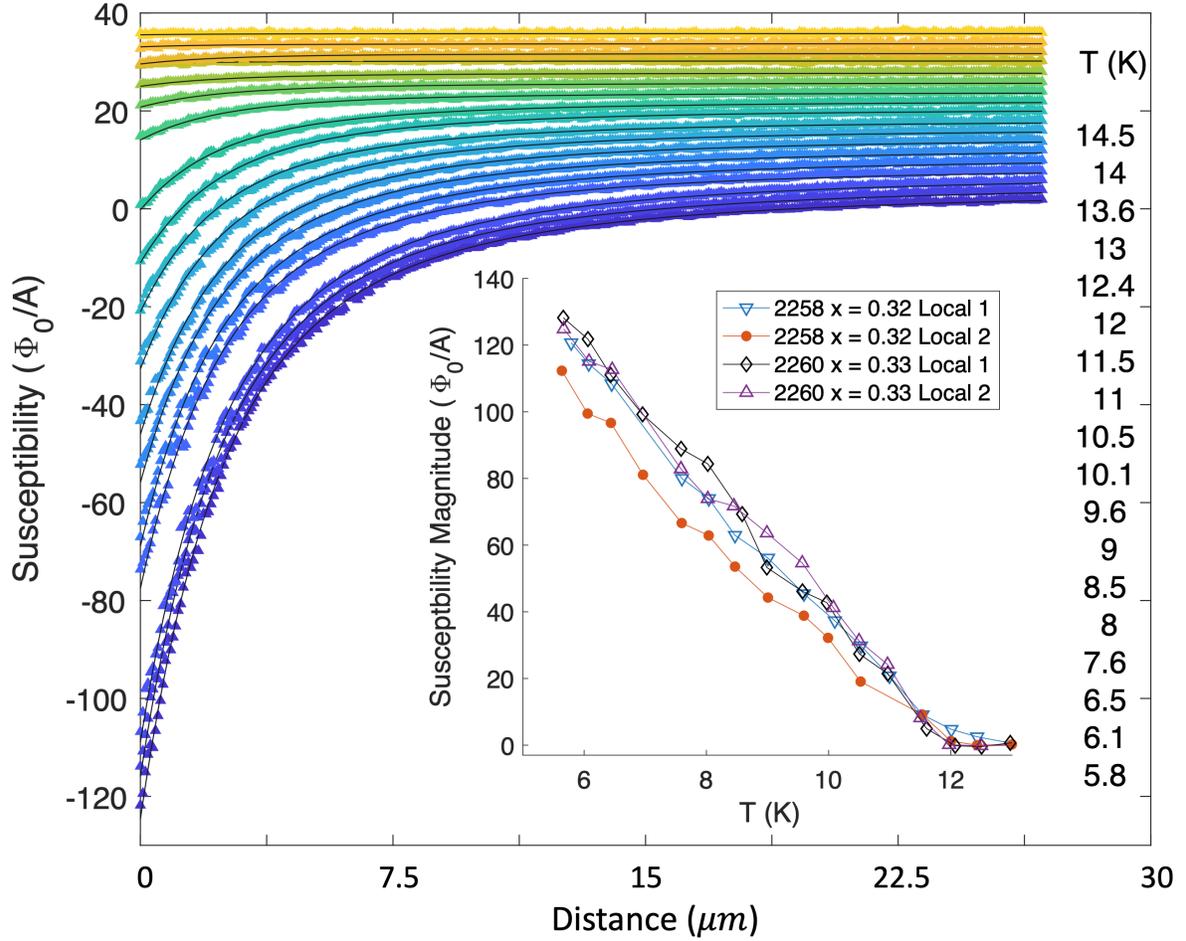

*FIG. 4. Typical susceptibility approach curves for the x = 0.32 LSCO film ('Local 1') as a function of voltage applied to the Z piezo reflecting the squid-sample distance. Each line corresponds to an approach curve at the set temperature. The largest-valued susceptibility approach curve is the one measured at the lowest temperature, 5.79 K. At higher temperatures above $T_c$, the curves flatten out to zero. The curves are offset vertically for clarity, nominally they all converge at zero for higher piezo voltage values. Inset: The overall change in the susceptibility from the furthest to the closest 'touchdown' distance from the sample. Note, the difference with scans values is due to difference in sample-SQUID separation distance.*

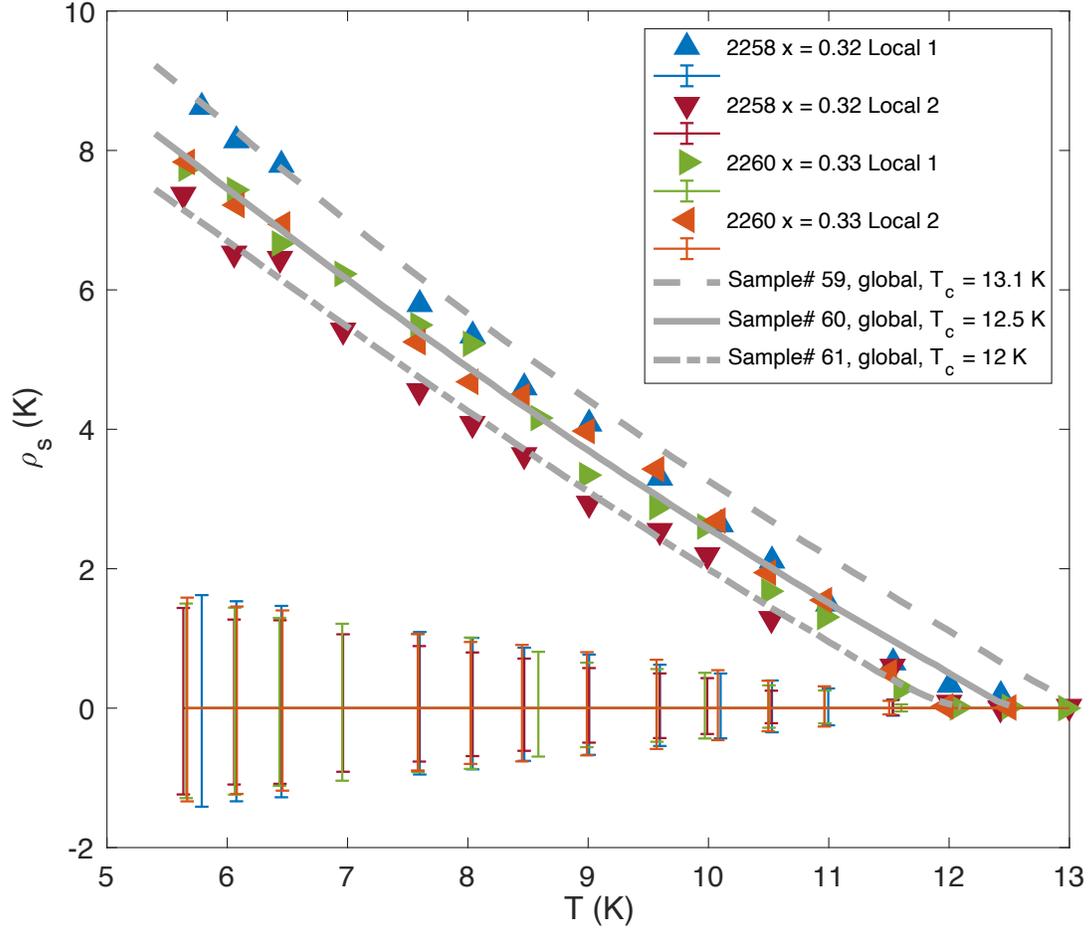

FIG. 5. *Local versus global superfluid stiffness. Triangles are the data based on the magnetic penetration depth extracted from the approach curves of the scanning SQUID sensor. Gray solid lines are experimental data from bulk measurements [4] (data for samples 59 – 61 are available on the publisher's website). Error bars represent possible variation in the temperature-slope of the local superfluid stiffness curves due to systematic uncertainties in $z_0$ of $\pm 0.5$ $\mu m$. The results show excellent agreement between global and local experiments.*

# Scanning SQUID characterization of extremely overdoped La₂₋ₓSrₓCuO₄

Chloe Herrera[1,2], Jacob Franklin[1], Ivan Božović[3,4], Xi He[3] and Ilya Sochnikov[1,5]

[1]Physics Department, University of Connecticut, Storrs CT 06269, USA
[2]Lux Research, Boston MA 02110, USA
[3]Brookhaven National Laboratory, Upton, New York 11973-5000, USA
[4]Department of Chemistry, Yale University, New Haven CT 06520, USA
[5]Institue of Material Science, University of Connecticut, Storrs CT 06269, USA

This Supplementary Material includes scanning SQUID susceptibility approach curves for the two samples and locations presented in the main text. Each plot of the approach curves is followed by a plot of the residuals of the fits. The data in each plot are shown by symbols color coded from dark blue to yellow, corresponding to increasing temperatures as indicated on the right. Solid lines represent fits to the model described in the main text. The data for each fixed-temperature approach curve are offset vertically for clarity.

In addition, we have included a plot of average values of susceptibility extracted from the scans shown in the main text. We emphasize that scans are less reliable for the evaluation of the superfluid density or the penetration depth because of larger variance in SQUID-to-sample distance than in the approach traces. The average values from the scans can be used to evaluate overall data consistency, but the results based on the approach traces shown in the main text are more accurate and appropriate for comparing to the bulk results.

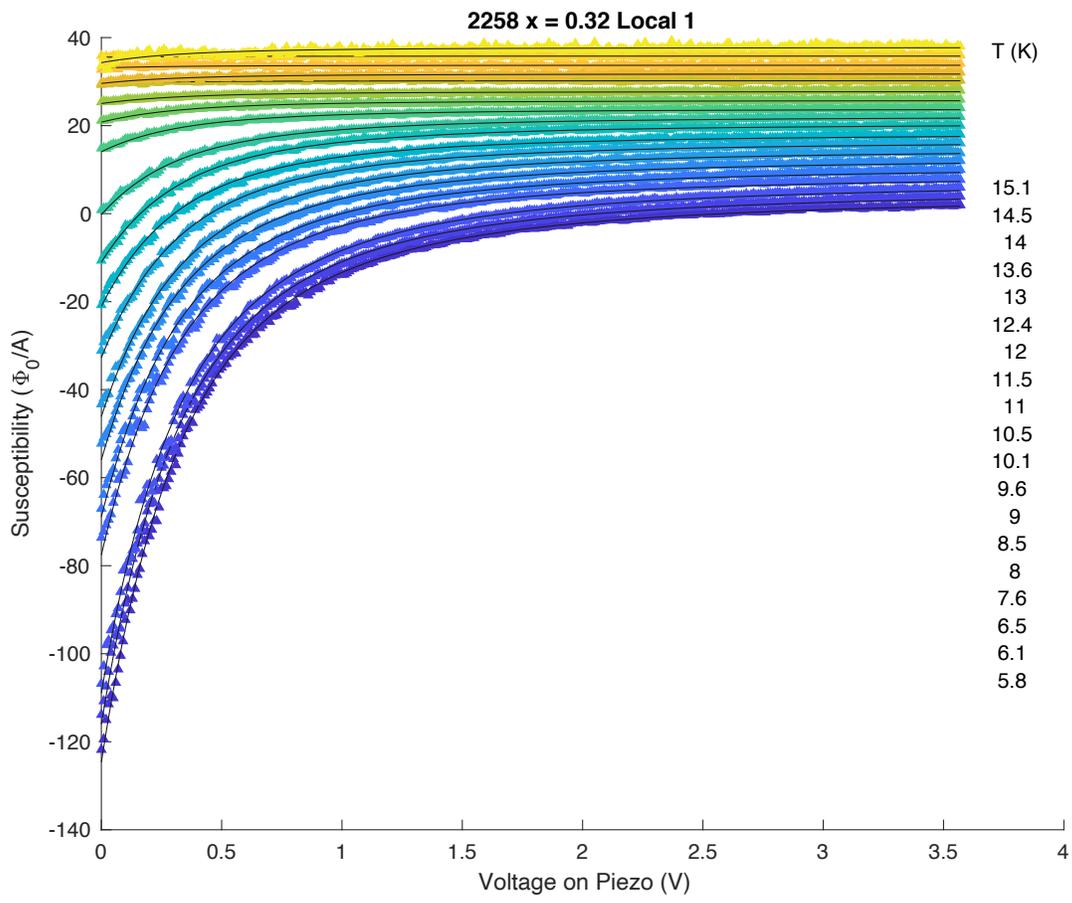

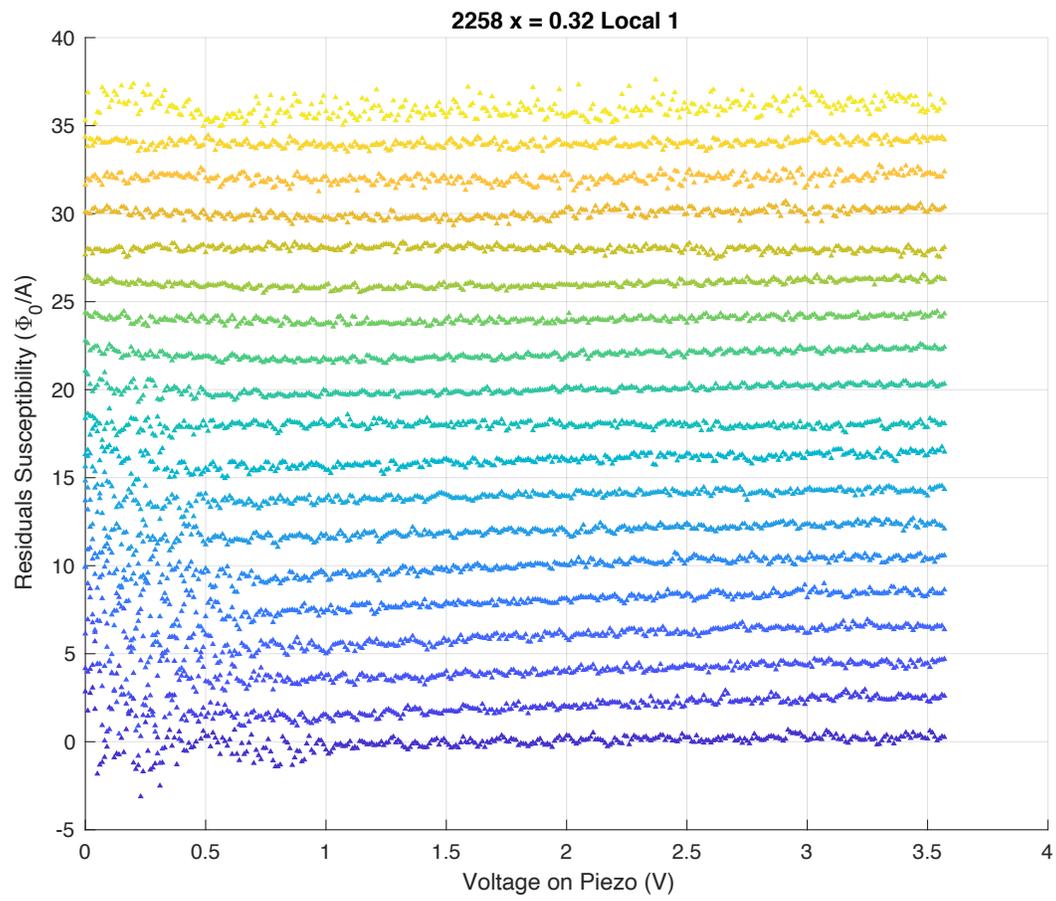

**2258 x = 0.32 Local 1**

Residuals Susceptibility ($\Phi_0$/A) vs Voltage on Piezo (V)

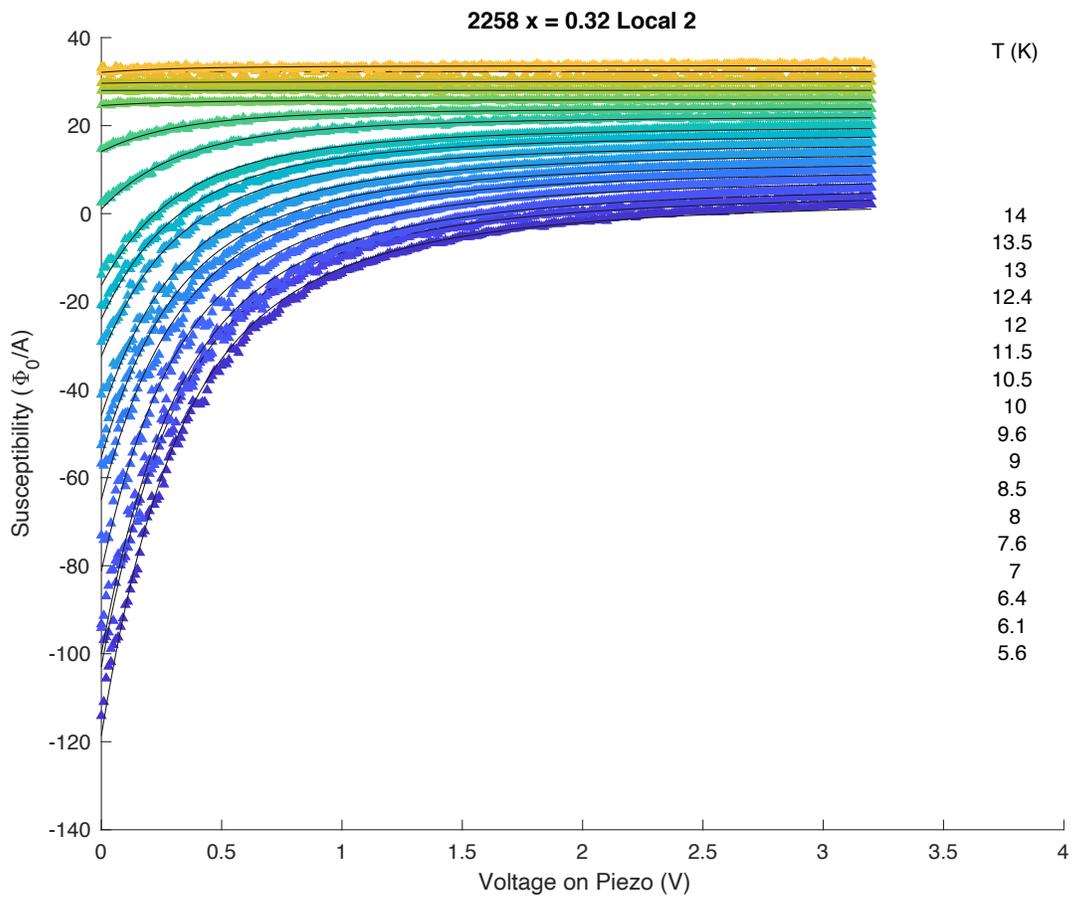

**2258 x = 0.32 Local 2**

T (K)

14
13.5
13
12.4
12
11.5
10.5
10
9.6
9
8.5
8
7.6
7
6.4
6.1
5.6

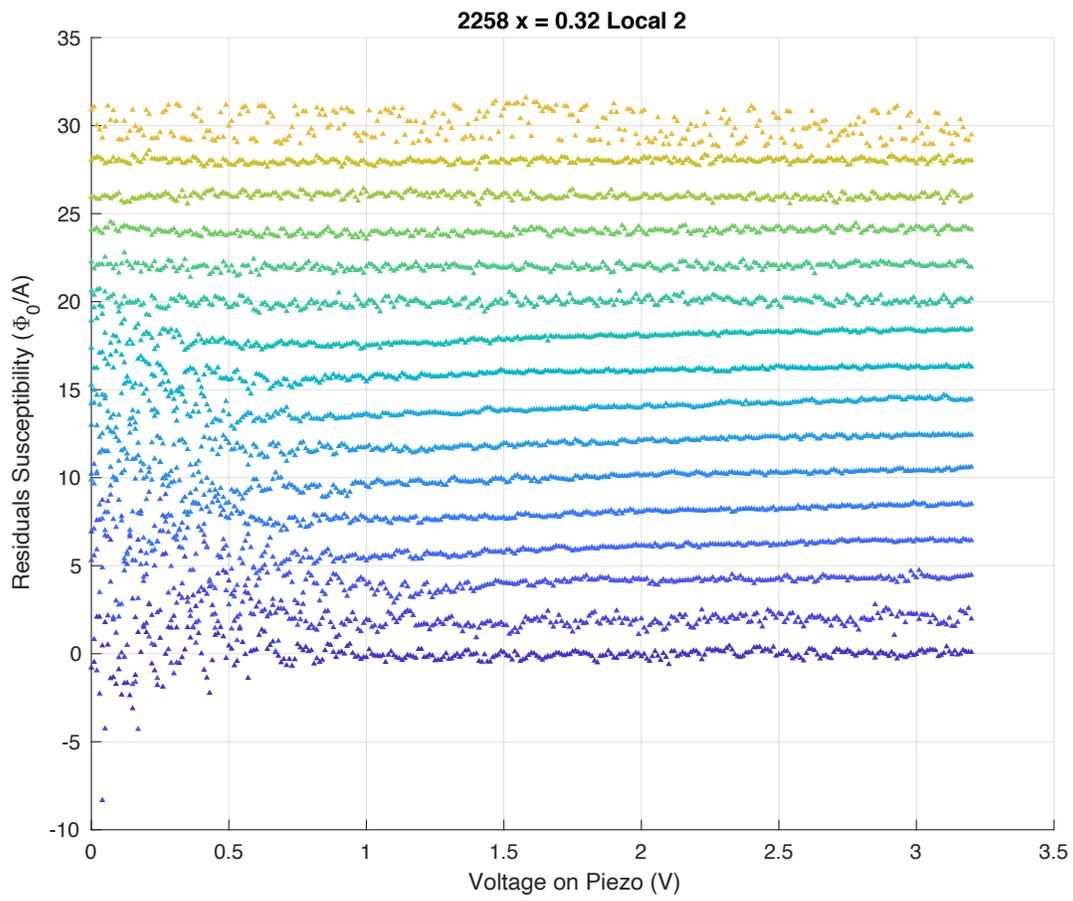

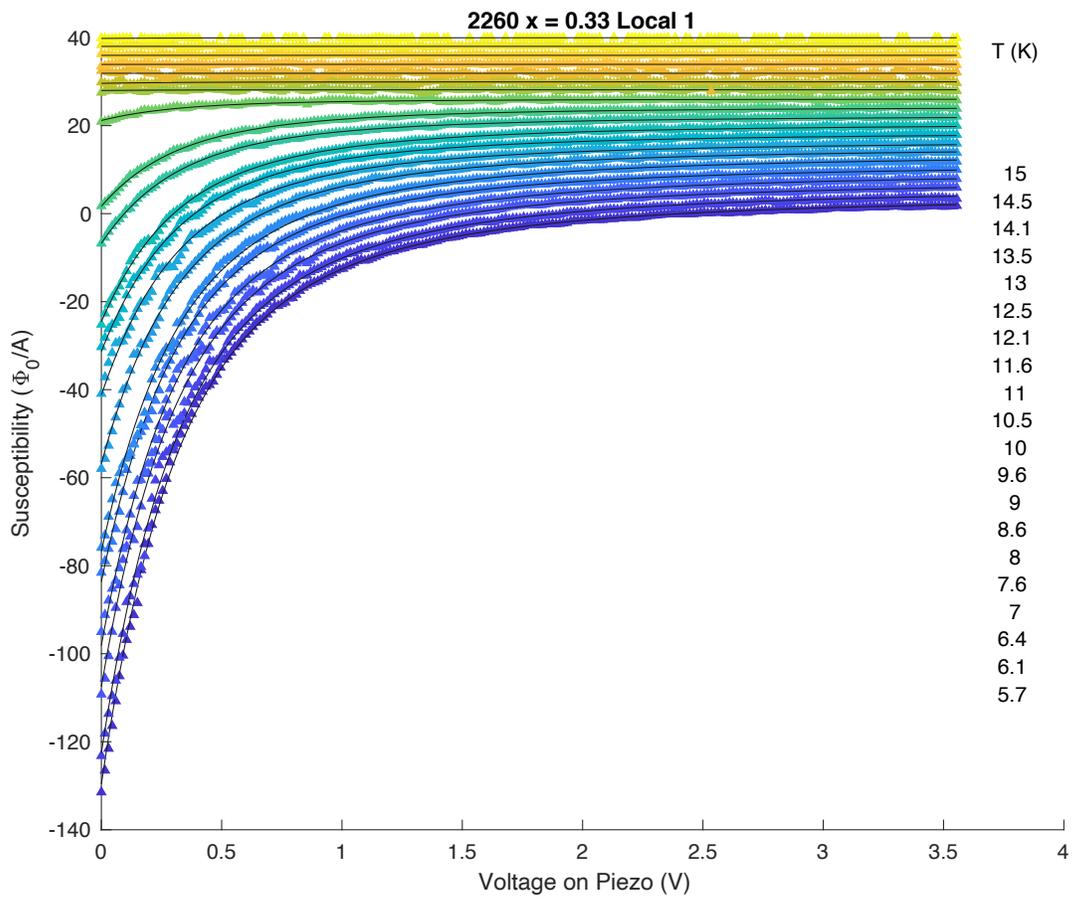

**2260 x = 0.33 Local 1**

T (K)

15
14.5
14.1
13.5
13
12.5
12.1
11.6
11
10.5
10
9.6
9
8.6
8
7.6
7
6.4
6.1
5.7

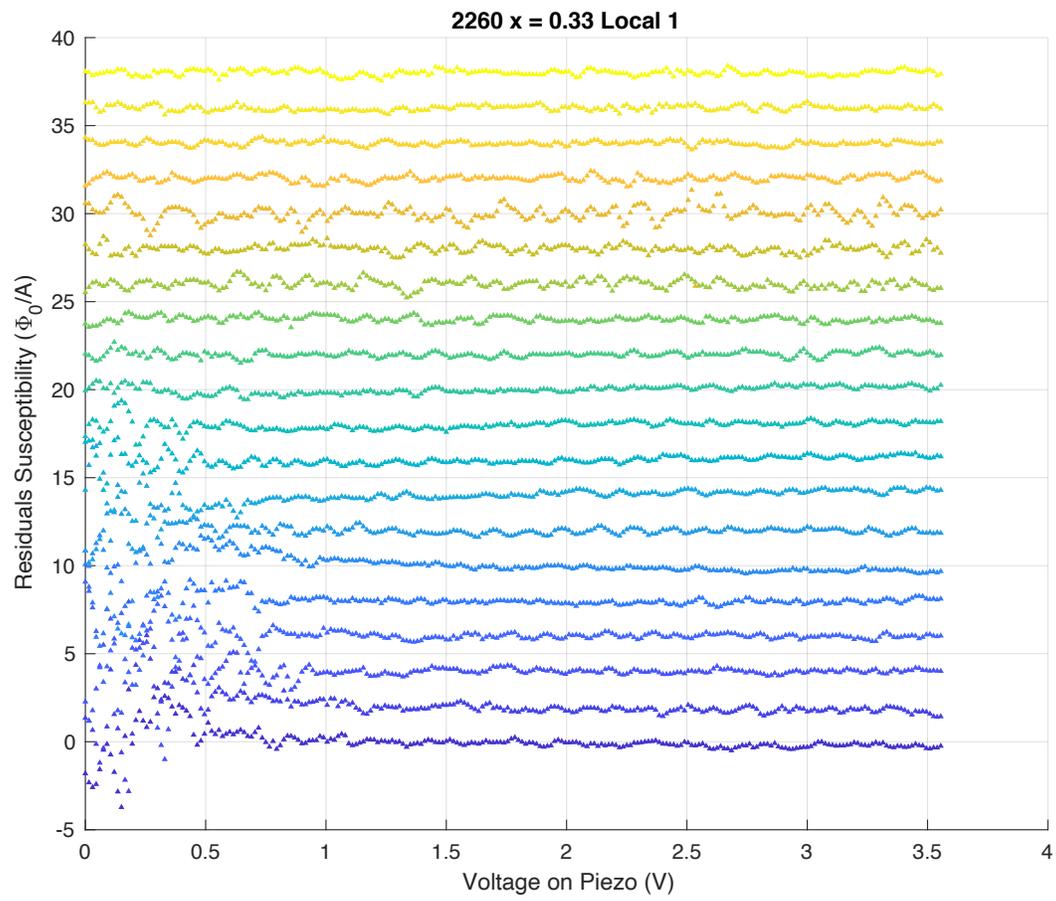

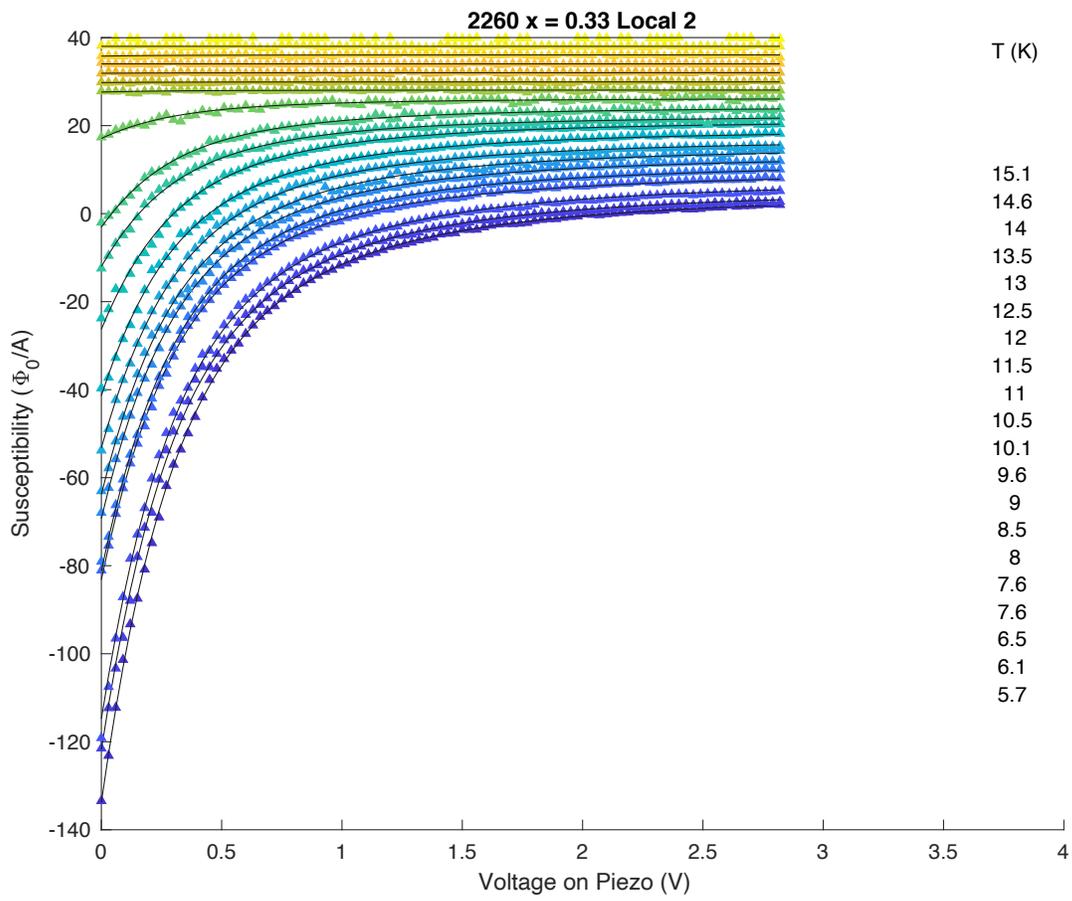

**2260 x = 0.33 Local 2**

T (K)

15.1
14.6
14
13.5
13
12.5
12
11.5
11
10.5
10.1
9.6
9
8.5
8
7.6
7.6
6.5
6.1
5.7

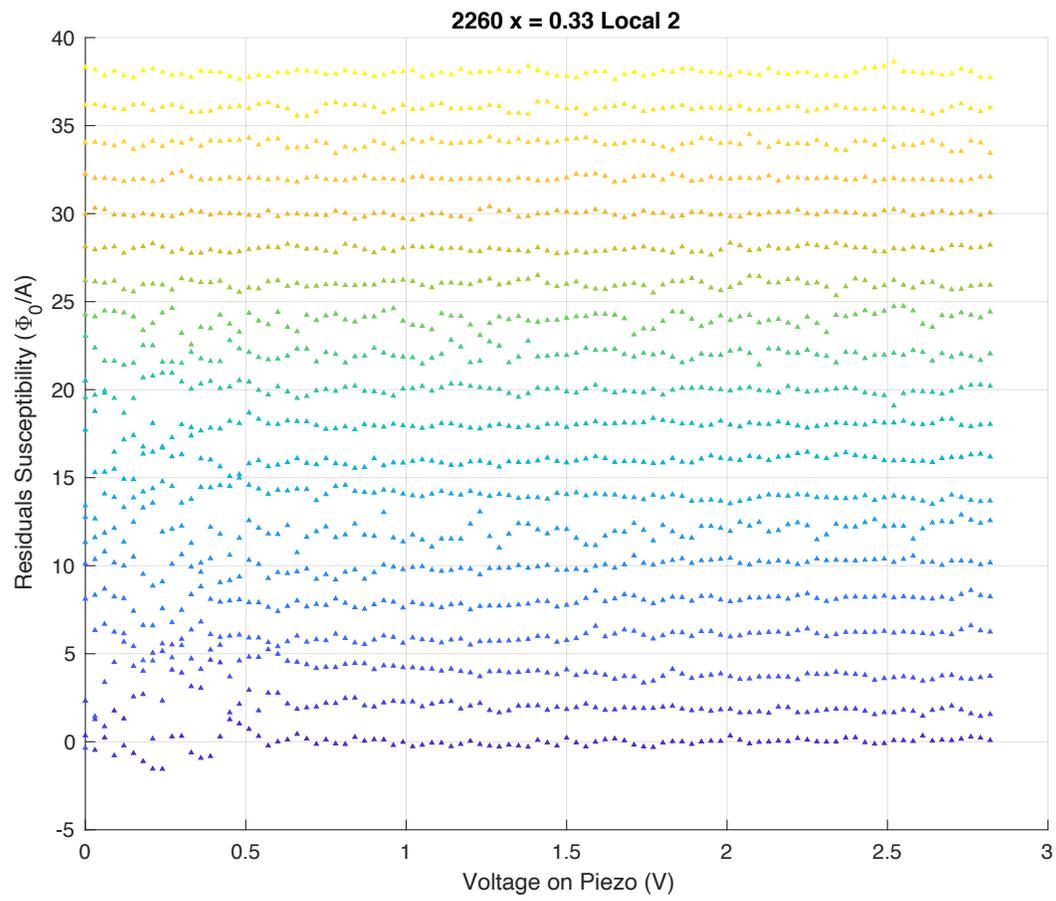

**2260 x = 0.33 Local 2**

Residuals Susceptibility ($\Phi_0$/A) vs Voltage on Piezo (V)

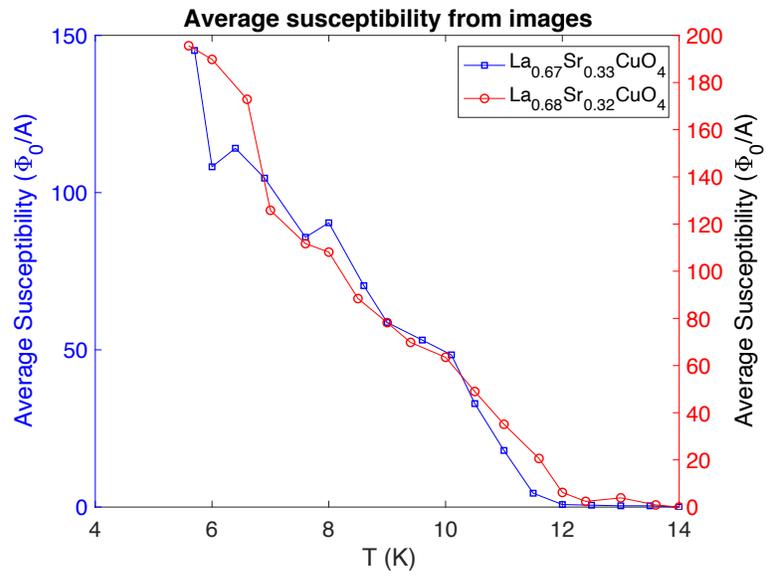